\pdfoutput=1
\documentclass[condensedmatter,article,accept,moreauthors,pdftex,10pt,a4paper]{mdpi}

\firstpage{1}
\pubvolume{3}
\issuenum{1}
\articlenumber{0}
\pubyear{2018}
\copyrightyear{2018}
\history{Received: 14 September 2018; Accepted: 26 November 2018; Published: date}
\updates{yes} % If there is an update available, un-comment this line

\usepackage{textcomp,bm}
\Title{The Magnetic Properties of 1111-type Diluted Magnetic Semiconductor (La$_{1-x}$Ba$_{x}$)(Zn$_{1-x}$Mn$_{x}$)AsO in the Low Doping Regime}

\Author{Guoxiang Zhi $^{1}$, Kai Wang $^{1}$, Haojie Zhang $^{1}$, Cui Ding $^{1}$, Shengli Guo $^{1}$, Yilun Gu $^{1}$, Licheng Fu $^{1}$ and F. L. Ning $^{1,2,}$*}%Please carefully check the accuracy of names and affiliations. Changes will not be possible after proofreading.

\AuthorNames{Guoxiang Zhi, Kai Wang, Haojie Zhang, Cui Ding, Shengli Guo, Yilun Gu, Licheng Fu, and F. L. Ning}

\address{%
$^{1}$ \quad Department of Physics, Zhejiang University, Hangzhou 310027, China; nucbill@zju.edu.cn (G.Z.); Kudoukai@zju.edu.cn (K.W.); havoc@zju.edu.cn (H.Z.); dingcui@zju.edu.cn (C.D.); slguo@zju.edu.cn (S.G.); 11636019@zju.edu.cn (Y.G.); 3120103824@zju.edu.cn (L.F.)\\ %Please confirm the author email is correct or not.
$^{2}$ \quad Collaborative Innovation Center of Advanced Microstructures, Nanjing University, Nanjing 210093, China}

\corres{Correspondence: ningfl@zju.edu.cn}

\abstract{We investigated the magnetic properties of (La$_{1-x}$Ba$_{x}$)(Zn$_{1-x}$Mn$_{x}$)AsO with $x$ varying from 0.005  to 0.05
at an external magnetic field of 1000 Oe. For doping levels of $x$ $\leq$ 0.01, the system remains paramagnetic down to the lowest measurable temperature of 2 K. Only when the doping level increases to $x$ = 0.02  does the ferromagnetic ordering appear. Our analysis indicates that antiferromagnetic exchange {\color{black}interactions dominate} for $x$ $\leq$ {\color{black}0.01}, as shown by the {\color{black}negative} Weiss temperature fitted from the magnetization data. The Weiss temperature becomes {\color{black}positive}, i.e., ferromagnetic coupling starts to dominate, for $x$ $\geq$ {\color{black}0.02}. The Mn--Mn spin interaction {\color{black}parameter $\mid$$2J/k_B$$\mid$ is} estimated to be {\color{black}in the order of 10 K for both $x$ $\leq$ 0.01 (antiferromagnetic ordered state) and $x$ $\geq$ 0.02 (ferromagnetic ordered state).} Our results unequivocally demonstrate the competition between ferromagnetic and antiferromagnetic exchange interactions in carrier-mediated ferromagnetic systems.}

\keyword{ferromagnetic ordering; antiferromagnetic exchange interaction; ferromagnetic coupling; carrier-mediated}

\begin{document}
%%%%%%%%%%%%%%%%%%%%%%%%%%%%%%%%%%%%%%%%%%
%% Only for the journal Gels: Please place the Experimental Section after the Conclusions

%%%%%%%%%%%%%%%%%%%%%%%%%%%%%%%%%%%%%%%%%%
%\setcounter{section}{-1} %% Remove this when starting to work on the template.
%\section{How to Use this Template}
%The template details the sections that can be used in a manuscript. Note that the order and names of article sections may differ from the requirements of the journal (e.g. the positioning of the Materials and Methods section). Please check the instructions for authors page of the journal to verify the correct order and names. For any questions, please contact the editorial office of the journal or support@mdpi.com. For LaTeX related questions please contact Janine Daum at latex-support@mdpi.com.
%The order of the section titles is: Introduction, Materials and Methods, Results, Discussion, Conclusions for these journals: aerospace,algorithms,antibodies,antioxidants,atmosphere,axioms,biomedicines,carbon,crystals,designs,diagnostics,environments,fermentation,fluids,forests,fractalfract,informatics,information,inventions,jfmk,jrfm,lubricants,neonatalscreening,neuroglia,particles,pharmaceutics,polymers,processes,technologies,viruses,vision

\section{Introduction}
Combining the properties of semiconductor and magnetism, diluted magnetic semiconductors (DMSs) have potential application in industry \cite{Furdyna, Zutic, Jungwirth}. In the 1980s, the focus was mainly on II-VI DMSs A$^{II}_{1-x}$Mn$_x$B$^{VI}$ (A = Zn, Cd, Hg; B = Se, Te), where iso-valent substitution of {\color{black}Mn$^{2+}$} for {\color{black}A$^{2+}$} in A$^{II}$B$^{VI}$ usually induces spin glass \cite{Furdyna}. The research of DMSs {\color{black}has become} explosive since III-V DMSs (In,Mn)As and (Ga,Mn)As were successfully fabricated via low temperature molecular beam epitaxy in the 1990s~\cite{InMnAs,GaMnAs,Chambers,Dietl1,Jungwirth,Samarth,Zutic,Dietl2}. It was initially proposed that the Curie temperature T$_C$ of III-V DMSs would reach room temperature assuming sufficient Mn ions could be doped homogeneously \cite{ZenerModel}. However, after more than two decades of effort, the highest T$_C$ of (Ga$_{1-x}$Mn$_x$)As has been reported as $\sim$200 K with $x$$\sim$12$\%$ \cite{GaMnAs_190K,ZhaoJH1,ZhaoJH2}. Some difficulties are encountered in the research of (Ga,Mn)As. Firstly, it is difficult to increase Mn concentration while keeping thin films homogeneous due to the low solid solubility of Mn in GaAs. Secondly, Mn substitution for GaAs provides not only local moments but also carriers. Moreover, some Mn$^{2+}$ ions behave as double donors when getting into interstitial sites \cite{Jungwirth}. Thirdly, it seems that T$_C$ does not {\color{black}increase } above 200 K where even higher Mn doping levels could be achieved, i.e., T$_C$ saturates at $\sim$200 K. The possible reason is that {\color{black}Mn$^{2+}$} are more likely to have another {\color{black}Mn$^{2+}$} at nearest-neighbour sites for higher Mn doping levels, which results in antiferromagnetic exchange interactions that {\color{black}compete with the} ferromagnetic ordering.

Recently, several bulk form DMS systems derived from Fe-based superconductors have been reported \cite{ZhaoJH2, Li(ZnMn)As_JCQ, Li(ZnMn)P_JCQ, Li(ZnCr)As_WQ, Li(ZnMn)P_Man, (BaK)(ZnMn)2As2_JCQ, ZhaoK_CSB_230K, (BaK)(CdMN)2As2_YXJ, (LaBa)(ZnMn)AsO_DC, (LaSr)(ZnMn)AsO_DC, (LaCa)(ZnMn)SbO_JCQ, (LaCa)(ZnMn)AsO_DC, (LaSr)(ZnFe)AsO_LJC, (LaSr)(CuMn)SO_YXJ}. Similar  to the style used in Fe-based superconductors, these DMS systems are named as ``111'' \cite{Li(ZnMn)As_JCQ,Li(ZnMn)P_JCQ,Li(ZnCr)As_WQ,Li(ZnMn)P_Man}, ``122'' \cite{(BaK)(ZnMn)2As2_JCQ,ZhaoK_CSB_230K,(BaK)(CdMN)2As2_YXJ}, ``1111'' \cite{(LaBa)(ZnMn)AsO_DC,(LaSr)(ZnMn)AsO_DC,(LaCa)(ZnMn)SbO_JCQ,(LaCa)(ZnMn)AsO_DC,
(LaSr)(ZnFe)AsO_LJC,(LaSr)(CuMn)SO_YXJ}, and ``32522'' \cite{32522_MHY} type DMSs. The progress on this research stream can be found in a recent review article \cite{Shengli_CPB}. The bulk form DMSs have some advantages. Firstly, we can apply conventional magnetic probes, such as muon spin relaxation ($\mu$SR) and nuclear magnetic resonance (NMR) to investigate the ferromagnetism at a microscopic level. $\mu$SR measurements have confirmed that the ferromagnetism in these DMSs is homogeneous and shares the same mechanism as that of (Ga,Mn)As \cite{Li(ZnMn)As_JCQ,Li(ZnMn)P_Man,(BaK)(ZnMn)2As2_JCQ,(LaBa)(ZnMn)AsO_DC,(LaCa)(ZnMn)AsO_DC,Dunsiger}. On the other hand, NMR results of Li(Zn$_{0.9}$Mn$_{0.1}$)P have demonstrated that Mn--Mn spin interactions are mediated by hole carriers, and the averaged ferromagnetic exchange {\color{black}interactions are} in the order of 100 K \cite{LiZnMnP_NMR}. Another advantage of these bulk form DMSs is that we can control the doping concentrations of spins and carriers separately, and their individual influence on ferromagnetic ordering can be investigated. For example, we found that there is no ferromagnetic order occurring when only doping Mn into the parent compound LaZnAsO up to 10$\%$ \cite{(LaBa)(ZnMn)AsO_DC}. The ferromagnetic order does not develop until carriers and spins are codoped into LaZnAsO. On the other hand, for a fixed amount of Mn concentration, too many carriers will suppress Curie temperature instead \cite{(LaSr)(ZnMn)AsO_DC,Li(ZnMn)P_Man}.

From theoretical point of view, {\color{black}the} NMR results on Li(Zn$_{0.9}$Mn$_{0.1}$)P are consistent with the theoretical model that {\color{black}spin interactions} are mediated by conduction carriers to form long range magnetic order. The long range ferromagnetic order has been explained by the p-d Zener exchange model \cite{ZenerModel}, which is an analog of the Ruderman--Kittel --Kasuya--%Please confirm whether there should be normal space between the ``-''
Yosida (RKKY) interaction in metals. However, in the very dilute regime, i.e., the concentrations of both spins and carriers are at a very low levels, carriers might be localized and cannot mediate the {\color{black}spin interactions}. Bound magnetic polaron (BMP) model, which proposes that the percolation of magnetic polarons will induce ferromagnetic ordering, has been applied to explain the mechanism in the very dilute regime of DMSs \cite{BMP}.

We   fully used the {\color{black}advantages} of decoupled charge and spin {\color{black}doping} in (La$_{1-x}$Ba$_{x}$)(Zn$_{1-x}$Mn$_{x}$)AsO, and investigated the magnetic properties in the {\color{black}very} dilute doped regime. We codoped Ba and Mn into the direct gap parent semiconductor LaZnAsO at low doping concentration{\color{black}, $x$ $\leq$ 0.05 
(more details about the synthesis and characterizations of (La$_{1-x}$Ba$_{x}$)(Zn$_{1-x}$Mn$_{x}$)AsO with higher doping level can be found in Ref. \cite{(LaBa)(ZnMn)AsO_DC}). }
We measured the magnetization at B$_{ext}$ = 1000 Oe and fit the data above T$_{C}$ to obtain the Weiss temperature $\theta$. Our fitting results indicate that weak antiferromagnetic exchange {\color{black}coupling} dominates for $x$ $\leq$ {\color{black}0.01}, and ferromagnetic coupling starts to dominate for $x$ $\geq$ {\color{black}0.02}. 
We {\color{black}calculated} the Mn--Mn spin interactions {\color{black}parameter $J$, and $2J/k_B$ is about $-$10 K $\pm$ 3 K for x $\leq$ 0.01 and about 17 K $\pm$ 5 K for x $\geq$ 0.02, which are in the same order as that of Hg$_{1-x}$Mn$_x$Te \cite{magnetic_IV-VI}.}
%We estimate the averaged Mn--Mn spin {\color{black}interactions} energy {\color{black}$\mid$$\frac{J}{k_B}$$\mid$ $\sim$ 1 K for $x$ $\leq$ 0.01 and $\mid$$\frac{J}{k_B}$$\mid$ $\sim$ 10 K for $x$ $\geq$ 0.02}, which are in the same order as that of {\color{black}Pb$_{1-x}$Mn$_x$Te and Hg$_{1-x}$Mn$_x$Te \cite{magnetic_IV-VI}}, respectively. 
We   show that the random nature of the spin {\color{black}interactions} for the doping regime of $x$ $\leq$ {\color{black}0.01} can be explained by {\color{black}models based on random-exchange interactions} \cite{HgMnTe,AndresPRB1981,MurayamaPRB1984,BhattPRL1982}.

%%%%%%%%%%%%%%%%%%%%%%%%%%%%%%%%%%%%%%%%%%
\section{Results and Discussion}

\begin{figure}[H]
\centering
\includegraphics[width=11 cm]{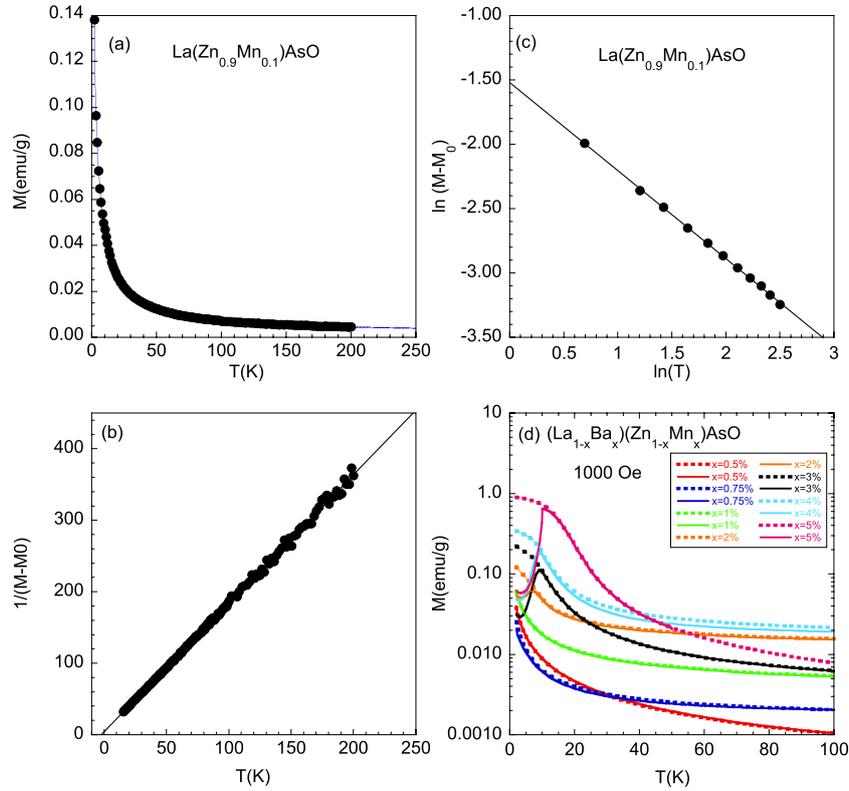}
\caption{(Color online) (\textbf{a}) Temperature dependence of magnetization
for LaZn$_{0.9}$Mn$_{0.1}$AsO under field cooling and zero field cooling
at B$_{ext}$ = 1000 Oe; the solid line represents {\color{black}the Curie--Weiss fit for temperature range of 15--200 K.}  %Please confirm whether there should be normal space between the ``-''
(\textbf{b}) Linear fit for {\color{black}1/(M $-$ M$_0$)} of LaZn$_{0.9}$Mn$_{0.1}$AsO for the temperature range of {\color{black}15--200 K, where M$_0$ is a temperature independent parameter derived from the Curie--Weiss fit.}  %Please confirm whether there should be normal space between the ``-''
(\textbf{c}) ln(M) versus ln(T) for LaZn$_{0.9}$Mn$_{0.1}$AsO. 
(\textbf{d}) Temperature dependence of magnetization for (La$_{1-x}$Ba$_x$)(Zn$_{1-x}$Mn$_{x}$)AsO ($x$ = 0.005$\sim$0.05) measured under field cooling {\color{black}(dashed lines)} and zero field cooling {\color{black}(solid lines)} at B$_{ext}$ = 1000 Oe.}
%\label{fig:M-T}
\label{fig1}
\end{figure}

In Figure \ref{fig1}a, we show the dc-magnetization of La(Zn$_{0.9}$Mn$_{0.1}$)AsO measured under field cooling (FC) and zero field cooling (ZFC) condition with B$_{ext}$ = 1000 Oe. The two curves are superposed to each other in the whole measured temperature range and can be well fitted by a Curie--Weiss relation \mbox{{\color{black}M = $\frac{C}{T-\theta}$+M$_0$}}, suggesting a paramagnetic ground state, where M$_{0}$ is temperature independent part induced by the host lattice (LaZnAsO).  
 {\color{black}(M $-$ M$_0$)$^{-1}$} versus T 
 {\color{black}(15 K to 200 K)} is shown in Figure \ref{fig1}b. The intercept in the T-axis is the Weiss temperature{\color{black}, $\theta$ = $-$2.6 K}, indicating the antiferromagnetic coupling. The negative intercept was also found for Hg$_{1-x}$Mn$_x$Te and Pb$_{1-x}$Mn$_x$Te, and other \mbox{II-VI~\cite{HgMnTe}} and IV-VI~\cite{magnetic_IV-VI} DMSs. 
 In Figure \ref{fig1}c, we show the plot of ln(M) versus ln(T), which can be fitted by a straight line $\mathrm{ln}(M-M_0) = -1.52 - 0.68\mathrm{ln}(T)$. This behavior indicates that the magnetization of La(Zn$_{0.9}$Mn$_{0.1}$)AsO follows a power-law dependence with temperature, with the slope {\color{black}$\alpha$ = $-$0.68.} The well {\color{black}fitted} results {\color{black}are} consistent with {\color{black}models based on random-exchange interactions} \cite{HgMnTe,AndresPRB1981,MurayamaPRB1984,BhattPRL1982}. {\color{black}In these models}, the {\color{black}low-temperature} magnetization in a system with random exchange {\color{black}interactions in any dimensions} follows a power-law with temperature as T$^{\alpha}$ ( $-$1 $\leq$ $\alpha$ $\leq$ 0 ) \cite{HgMnTe}. The magnitude of $\alpha$ is a measure of how random the spin interactions are. For example, $\alpha$ = $-$1 corresponds to the standard Curie-law behavior \cite{HgMnTe}.

We show the field cooling (FC) and zero field cooling (ZFC) dc-magnetization of (La$_{1-x}$Ba$_{x}$)(Zn$_{1-x}$Mn$_{x}$)AsO ($x$ = 0.005$\sim$0.05) for B$_{ext}$ = 1000 Oe in Figure \ref{fig1}d. 
The magnitude of magnetization at low temperature (2 K) measured under FC condition increases from \mbox{{0.039} emu/g (x = 0.005)} to 0.903 emu/g (x = 0.05). 
For $x$ $\leq$ 0.01, no   magnetic instability or anomaly is observed in both FC and ZFC curves, indicating there is no   ferromagnetic or spin-glass ordering induced. 
Starting from $x$ = 0.02, a clear splitting between FC and ZFC curves at low temperature can be observed. 
We fit the data of susceptibility (M/H) at high enough {\color{black}(much higher than $\theta$)} temperature by a Curie--Weiss law {\color{black}$\chi$ = $\frac{C}{T - \theta}$+$\chi_0$}, where {\color{black} $\chi_0$ is temperature independent part induced by the host lattice,} C is the Curie constant and $\theta$ the Weiss temperature. 
With the number C, we can calculate {\color{black}the effective occupation probability of magnetic ion on a Zn site, $\bar{x}$. 
Considering a modified Brillouin function with Weiss' molecule field theory, the magnetization of $\bar{x}N_0$ Mn ions is $M=S_{Mn}g_{Mn}\mu_{B}\bar{x}N_0B_{S_{Mn}}(\xi)$, where $ \xi=\frac{g_{Mn}\mu_{B}S_{Mn}H_{eff}}{k_{B}T} $, $ H_{eff}=H_e+\lambda M $, $S_{Mn}$ is the spin of Mn, g$_{Mn}$ is the spin-splitting g factor of Mn, $\mu_B$ is the Bohr magneton, $N_0$ is the number of Zn site per gram, $k_B$ is the Boltzmann constant, $H_{eff}$ is the effective magnetic field, $H_e$ is the applied field and $\lambda$ is a constant. 
Since the Brillouin function $B_{S}(\xi)=\frac{S+1}{3S}\xi + O(x^{3})$, we have $ \chi=\frac{\bar{x}N_0(g_{Mn}\mu_{B})^{2}S_{Mn}(S_{Mn}+1)}{3k_B(T-\theta')} $, where $ \theta'= \frac{\bar{x}N_0(g_{Mn}\mu_{B})^{2}S_{Mn}(S_{Mn}+1)\lambda}{3k_B}$. 
Comparing with the Curie--Weiss law, we have $\theta'$=$\theta$ and} $\bar{x}$ =$\frac{3k_{B}CM_{mol}}{(g_{Mn}\mu_{B})^2S_{Mn}(S_{Mn}+1)N_A}$, where M$_{mol}$ is the formula weight and N$_A$ is the Avogadro number. 
We assume that g$_{Mn}$ = 2 and S$_{Mn}$ = $\frac{5}{2}$ ({\color{black}this assumption is reasonable because the effective moment from experiments is estimated to be \mbox{4--5 $\mu_B$/Mn~\cite{(LaBa)(ZnMn)AsO_DC}}}). 

{\color{black}From the Heisenberg model, if we count only the nearest-neighbor exchange interaction of Mn ions, we can derive the molecular field $H_{mf}=\lambda M=\frac{1}{g\mu_B}\sum\limits_{\textbf{R}'}\bar{x}J(\textbf{R}-\textbf{R}')S(\textbf{R}')$. 
Assuming every spin has the same value and sums up to give the total magnetization, we have $S(\textbf{R}')$  = $\frac{M}{g\mu_B\bar{x}N_0}$, and $\lambda=\frac{2\bar{x}zJ}{\bar{x}N_0(g\mu_B)^{2}}$. 
Under the circumstance that temperature is in the range where no long-range magnetic order formed,} the nearest--neighbor exchange interaction can be estimated by using $\theta$ and $\bar{x}$ from the relation {\color{black}$J$ = $\frac{ 3{\theta}k_B}{2\bar{x}S_{Mn}(S_{Mn}+1)z}$}, where $z$ is the number of nearest neighbors on Zn site. 
{\color{black}$z$ equals} to 4 in (La$_{1-x}$Ba$_{x}$)(Zn$_{1-x}$Mn$_{x}$)AsO system.

In Table \ref{tab1}, we list the concentration $x$, obtained {\color{black}$\bar{x}$,} $\theta$ and $J$ for (La$_{1-x}$Ba$_{x}$)(Zn$_{1-x}$Mn$_{x}$)AsO. 
%The parameters for $x$ $>$ 0.05 are estimated by the magnetization data shown in ref. 21. 
The small {\color{black}negative} Weiss temperature $\theta$ for $x$ $\leq$ {\color{black}0.01} indicates that weak antiferromagnetic interactions dominate. On the other hand, starting from x = {\color{black}0.02}, the Weiss temperature $\theta$ turns into a {\color{black}positive} value, {\color{black}indicating} ferromagnetic
{\color{black}interactions start} to dominate. Note that, in Pb$_{1-x}$Eu$_x$Te, the coupling between Eu ions is still antiferromagnetic for the doping level up to 31.6 \% \cite{magnetic_IV-VI}.

\begin{table}[H]
\caption{Weiss temperature ($\theta$) and the nearest neighbor exchange interaction for (La$_{1-x}$Ba$_{x}$)(Zn$_{1-x}$Mn$_{x}$)AsO.}\label{tab1}
\centering
%% \tablesize{} %% You can specify the fontsize here, e.g.  \tablesize{\footnotesize}. If commented out \small will be used.
\begin{tabular}{cccc}
\toprule
\textbf{x} & $\bar{\textbf{x}}$ & \textbf{{\bm{$\theta$}} (K)}	& \textbf{2{\bm{$J$}}/k$_B$ (K)}\\
\midrule
       0.005  &   0.0064 $\pm$ 0.0003  & $-$0.9 $\pm$ 0.1  &  $-$11.71 \\
       0.0075 &   0.0034 $\pm$ 0.0002  & $-$0.3 $\pm$ 0.1  &   $-$7.47 \\
       0.01   &   0.011 $\pm$ 0.001    & $-$1.7 $\pm$ 0.2  &  $-$13.45 \\
       0.02   &   0.018 $\pm$ 0.001    &  2.5 $\pm$ 0.2  &   12.12 \\
       0.03   &   0.032 $\pm$ 0.002    &  5.8 $\pm$ 0.1  &   15.44 \\
       0.04   &   0.046 $\pm$ 0.003    &  6.2 $\pm$ 0.1  &   11.64 \\
       0.05   &   0.066 $\pm$ 0.003    & 16.2 $\pm$ 0.2  &   21.22 \\
%       0.10   &   0.074$\pm$0.004    & 25.1$\pm$0.1  &   29.15 \\
%       0.15   &   0.10$\pm$0.01      & 24.7$\pm$0.2  &   20.76 \\
%       0.20   &   0.065$\pm$0.003    & 24.2$\pm$0.2  &   31.84 \\
\bottomrule
\end{tabular}
\end{table}

\begin{figure}[H]
\centering
\includegraphics[width=11 cm]{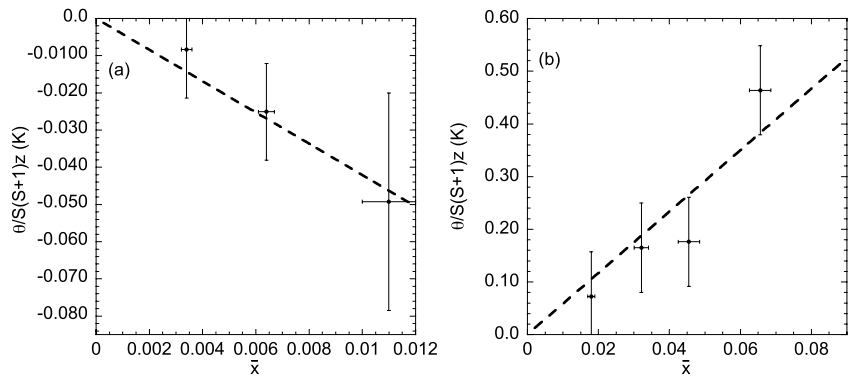}
\caption{{The parameter $\frac{\theta}{S_{Mn}(S_{Mn}+1)z}$} versus the effective magnetic ion concentration
 $\bar{x}$ in (La$_{1-x}$Ba$_x$)(Zn$_{1-x}$Mn$_x$)AsO for: (\textbf{a}) 0.005 $\leq$ $x$ $\leq$ {\color{black}0.01}; and (\textbf{b}) {\color{black}0.02} $\leq$ $x$ $\leq$ {\color{black}0.05}. {\color{black}Dashed} lines {\color{black}are guides for eyes.}}
%\label{fig:fitJ}
\label{fig2}
\end{figure}

%Usually we can also estimate the averaged $\mid$ $J$ $\mid$ through the analysis of the parameter {$\frac{\theta}{S_{Mn}(S_{Mn}+1)z}$} versus the effective magnetic ion concentration $\bar{x}$. 
In the first approximation, {$\frac{\theta}{S_{Mn}(S_{Mn}+1)z}$} should depend linearly on $\bar{x}$ and go to zero when $\bar{x}$ goes to zero, with a slope of $\frac{2J}{3k_B}$ \cite{magnetic_IV-VI}. We show the plot of {$\frac{\theta}{S_{Mn}(S_{Mn}+1)z}$} versus $\bar{x}$ of {(La$_{1-x}$Ba$_x$)(Zn$_{1-x}$Mn$_x$)AsO} for 0.005 $\leq$ $x$ $\leq$ {\color{black}0.01} and {\color{black}0.02} $\leq$ $x$ $\leq$ {\color{black}0.05} in Figure \ref{fig2}a,b, respectively. 
{\color{black}The {\color{black}dashed} linear lines passing through the origin are guides for eyes. In Figure \ref{fig2}, the data points are not far away from a linear line, which gives   evidence for the validity of calculating $\bar{x}$ with such method.}  
%Forcing the linear fitting lines pass through the origin, we obtain the average $\mid\frac{J}{k_B}\mid$ $\sim$ {\color{black}6.71} K and {\color{black}11.26} K for 0.005 $\leq$ $x$ $\leq$ {\color{black}0.01} and {\color{black}0.02} $\leq$ $x$ $\leq$ 0.2, respectively. 

For $x\geq 0.02$, the nearest-neighbor interaction parameter $\mid2J/k_B\mid$ of (La$_{1-x}$Ba$_x$)(Zn$_{1-x}$Mn$_x$)AsO with ferromagnetic {\color{black}coupling} is {\color{black}about 17 K $\pm$ 5 K.}
{\color{black} For x $\leq$ 0.01, with antiferromagnetic coupling, $\mid2J/k_B\mid$} is {\color{black}about 10 K $\pm$ 3 K}, which is {\color{black}in the same order of magnitude as that in} Mn-doped II-VI systems \cite{magnetic_IV-VI}. For instance, $\mid2J/k_B\mid$ for Mn-doped HgSe, CdSe and CdTe are 21.8 K, 21.2 K and 13.8 K, respectively \cite{HgCdMnSeCdMnTe}. 
However, this value is much larger than {\color{black}that in} IV-VI systems with {\color{black}Mn or rare-earth doped.} 
{\color{black}$\mid2J/k_B\mid$} for Mn-doped IV-VI compounds is in the order of 1 K \cite{magnetic_IV-VI}{\color{black}. F}or example, $\mid2J/k_B\mid$ for {\color{black}Mn-doped} PbS, PbSe and PbTe are 2.57 K \cite{PbMnS}, 3.35 K \cite{magnetic_IV-VI} and 1.69 K \cite{magnetic_IV-VI}, respectively.
{\color{black}F}or rare-earth-doped IV-VI compounds, {\color{black}$\mid2J/k_B\mid$} is in the order of 0.1 K \cite{magnetic_IV-VI}{\color{black}. F}or instance, {\color{black}$\mid2J/k_B\mid$} for {Pb$_{1-x}$Eu$_x$Te} \cite{PbEuTe} and {Pb$_{1-x}$Gd$_x$Te} \cite{magnetic_IV-VI} {\color{black}is} 0.23 K and 0.72 K, respectively. This means that  the {\color{black}magnitude} of nearest-neighbor exchange interaction in $(La_{1-x}Ba_x)(Zn_{1-x}Mn_x)AsO $ is an order of magnitude higher than that in Mn-doped IV-VI systems, and two orders of magnitude higher than that in rare-earth-doped IV-VI systems.

The magnitude of the nearest-neighbor exchange interaction in (La$_{1-x}$Ba$_x$)(Zn$_{1-x}$Mn$_x$)AsO is quite different from that in Mn-doped IV-VI systems; one possible reason is that $J$ strongly depends on the cation-anion distance in DMSs \cite{magnetic_IV-VI}. CdTe(II-VI DMS), HgTe(II-VI DMS) and PbTe(IV-VI DMS) share almost the same lattice constant $a$ ($\sim$6.5 \AA), but their magnitude of exchange interactions are quite different. The cation-anion separation of II-VI systems with zinc-blende or wurtzite structure can be calculated as $a\sqrt{3}/4$ ($\sim$2.8 \AA), while the cation-anion separation of IV-VI systems with rock salt structure is $0.5a$ ($\sim$3.3 \AA). M. Gorska and J. R. Anderson calculated that this difference in the cation-anion separation alone can lead to an order of magnitude or more difference in the superexchange interaction parameter $J$ \cite{magnetic_IV-VI}, and they believe that superexchange is the dominant exchange mechanism for these DMSs. For LaZnAsO with a ZrCuSiAs-type tetragonal crystal structure, the cation-anion separation of Mn-doped ZnO layer is $\sim$ 2.8 \AA, which is close to that in II-VI systems, and explains why the value of J  in (La$_{1-x}$Ba$_x$)(Zn$_{1-x}$Mn$_x$)AsO $(x \leq 0.01)$ is higher than that in IV-VI systems and close to that in II-VI systems.

\begin{figure}[H]
\centering
\includegraphics[width=13 cm]{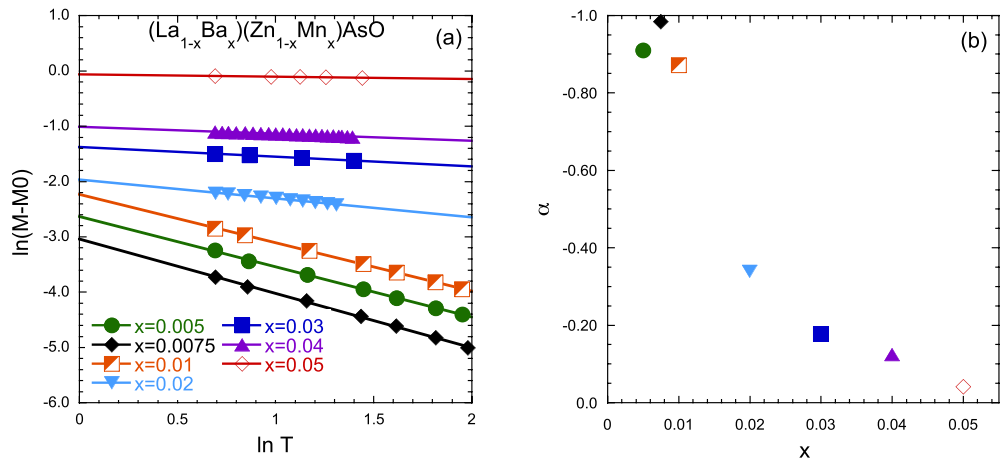}
\caption{{\color{black}(Color online) }(\textbf{a}) {\color{black}ln(M $-$ M$_0$)} versus ln(T) for (La$_{1-x}$Ba$_x$)(Zn$_{1-x}$Mn$_x$)AsO
(0.005 $\leq$ $x$ $\leq$ {\color{black}0.05}). Solid lines represent the linear fit.
(\textbf{b}) $\alpha$ obtained in (\textbf{a}) versus the doping concentration $x$.}
\label{lnMlnT}
\end{figure}

We have seen that in, the very dilute regime {\color{black}($x$ $\leq$ 0.01)}, even carriers and spins are codoped; (La$_{1-x}$Ba$_x$)(Zn$_{1-x}$Mn$_x$)AsO do not show any ferromagnetic or spin-glass ordering, similarly to the case of La(Zn$_{0.9}$Mn$_{0.1}$)AsO (where no carriers are introduced). This behavior can be explained by {\color{black}models based on random exchange interactions, in which} the low-temperature susceptibility follows a power law in temperature as $T^{\alpha}$ with $-1 \leq \alpha \leq 0$ \cite{HgMnTe}. 
We plot ln(M$-$M$_0$) versus ln(T) of (La$_{1-x}$Ba$_x$)(Zn$_{1-x}$Mn$_x$)AsO for 0.005 $\leq$ $x$ $\leq$ 0.02 in Figure \ref{lnMlnT}a. The data can be fitted by  {\color{black}straight lines} with the slopes as the random-exchange {\color{black}parameters, $\alpha$}. 
The plot of $\alpha$ versus concentration $x$ is shown in Figure \ref{lnMlnT}b. 
Note that $\alpha$ {\color{black}equal } to $-$1 corresponds to a standard Curie-law behavior. For $x$ = 0.005, $\alpha$ is $-$0.91, which is  {\color{black}quite} close to a fully paramagnetic ground state. 
{\color{black}For $x$ = 0.0075, $\alpha$($\sim$$-$0.98) is closer to $-$1 than that for $x$ = 0.005, which is consistent with the magnitude of effective occupation probability of magnetic ion, $\bar{x}$($\sim$0.0034 for x = 0.0075, which is smaller than $\bar{x}$$\sim$0.0064 for x = 0.005).} As   more Ba and Mn are doped, the absolute value of $\alpha$ is suppressed quickly. For $x$ = 0.02, the absolute value of $\alpha$ already {\color{black}decreases} to $\sim$$-$0.3, as shown in Figure \ref{lnMlnT}b. The suppression of magnitude of $\mid\alpha\mid$ with increasing $x$ indicates that increasing carriers and spins concentration enhances the coupling between Mn ions, and eventually results in the formation of ferromagnetic long range ordering.

%%%%%%%%%%%%%%%%%%%%%%%%%%%%%%%%%%%%%%%%%%
%\section{Discussion}

%Authors should discuss the results and how they can be interpreted in perspective of previous studies and of the working hypotheses. The findings and their implications should be discussed in the broadest context possible. Future research directions may also be highlighted.

%%%%%%%%%%%%%%%%%%%%%%%%%%%%%%%%%%%%%%%%%%
\section{Materials and Methods}

(La$_{1-x}$Ba$_{x}$)(Zn$_{1-x}$Mn$_{x}$)AsO polycrystalline specimens were synthesized by solid-state reaction method as   described in \cite{(LaBa)(ZnMn)AsO_DC}. High purity elements of La, Zn and As were mixed and heated slowly to 900 $^{\circ}\mbox{C}$ in evacuated silica tubes, and held for 15 h before shutting off the furnace to produce intermediate products LaAs and ZnAs. LaAs and ZnAs were then mixed with ZnO, BaO$_{2}$ and Mn with the stoichiometric composition. The mixture was pressed into pellets and slowly heated up to 1150 $^{\circ}\mbox{C}$ and kept for 40 h before cooling at a rate of 10 $^{\circ}\mbox{C}$/h to room temperature. The polycrystals were characterized via X-ray diffraction at room temperature using a PANalytical X-ray diffractometer (Model EMPYREAN) with monochromatic $\mbox{CuK}_{\alpha1}$ radiation. The dc-magnetization measurements were conducted on a Quantum Design superconducting quantum interference device (SQUID).

%%%%%%%%%%%%%%%%%%%%%%%%%%%%%%%%%%%%%%%%%%
\section{Conclusions}

The carrier and spin density of (La,Ba)(Zn,Mn)AsO can be accurately controlled with the doping level ranging from 0.005 to {\color{black}0.05}. No ferromagnetic ordering occurs when only doping Mn into the parent compound LaZnAsO up to 10\%. With both carriers and moments {\color{black}doping}, ferromagnetic ordering starts to form when x $\geq$ 0.02. {\color{black}With} the doping level increasing, the obtained Weiss temperature $\theta$ changes {\color{black}from negative to positive at $x$ = 0.02}, indicating the {\color{black}dominated} exchange {\color{black}coupling} between Mn ions is transformed from antiferromagnetic to ferromagnetic. For {\color{black}$x$ $\leq$ 0.02}, the magnetization at low temperature is proportional to T$^\alpha$ with $-1\leq\alpha\leq 0$, which can be explained by {\color{black}models based on random-exchange interactions}. The nearest neighbor exchange interaction {\color{black}parameter $\mid2J/k_B\mid$} is {\color{black}about 10 K $\pm$ 3 K and 17 K $\pm$ 5 K for x $\leq$ 0.01 and $x\geq 0.02$},  respectively. Possessing the {\color{black}advantages} of decoupled spin and charge doping, (La,A)(Zn,B)PnO (A = Ca, Sr, Ba; B = Mn, Fe, Pn = As, P) are chemically stable systems for investigation of ferromagnetic mechanism \cite{(LaBa)(ZnMn)AsO_DC,(LaSr)(ZnMn)AsO_DC,(LaCa)(ZnMn)SbO_JCQ,
(LaSr)(ZnFe)AsO_LJC}.

%%%%%%%%%%%%%%%%%%%%%%%%%%%%%%%%%%%%%%%%%%
%\section{Patents}
%This section is not mandatory, but may be added if there are patents resulting from the work reported in this manuscript.

%%%%%%%%%%%%%%%%%%%%%%%%%%%%%%%%%%%%%%%%%%
\vspace{6pt}

%%%%%%%%%%%%%%%%%%%%%%%%%%%%%%%%%%%%%%%%%%
%% optional
%\supplementary{The following are available online at \linksupplementary{s1}, Figure S1: title, Table S1: title, Video S1: title.}

% \supplementary{The following are available at \linksupplementary, Figure S1: title, Table S1: title, Video S1: title. A supporting video article is available at doi: link.}

%%%%%%%%%%%%%%%%%%%%%%%%%%%%%%%%%%%%%%%%%%
\authorcontributions{Conceptualization, F.L.N.; formal analysis, G.Z. and C.D.; investigation, C.D., K.W., H.Z., S.G., G.Z., Y.G. and L.F.; writing---original draft preparation, C.D.; writing---review and editing, G.Z. and F.L.N.; visualization, C.D. and G.Z.; and supervision, F.L.N.}

%%%%%%%%%%%%%%%%%%%%%%%%%%%%%%%%%%%%%%%%%%
\funding{This research was funded by Key Projects for Research and Development of China (No. 2016FYA0300402), National Natural Science Foundation of China (No.11574265), Natural Science Foundation of Zhejiang Province (No. LR15A040001), and the Fundamental Research Funds for the Central Universities (No. 2017FZA3003).}

%%%%%%%%%%%%%%%%%%%%%%%%%%%%%%%%%%%%%%%%%%
\acknowledgments{F.L. Ning acknowledges helpful discussions with R.N. Bhatt.}

%%%%%%%%%%%%%%%%%%%%%%%%%%%%%%%%%%%%%%%%%%
\conflictsofinterest{The authors declare no conflict of interest. The founding sponsors had no role in the design of the study; in the collection, analyses, or interpretation of data; in the writing of the manuscript, or in the decision to publish the results.}

%%%%%%%%%%%%%%%%%%%%%%%%%%%%%%%%%%%%%%%%%%
%% optional
\abbreviations{The following abbreviations are used in this manuscript:\\

\noindent
\begin{tabular}{@{}ll}
DMSs & Diluted Magnetic Semiconductors\\
$\mu$SR & Muon Spin Relaxation\\
NMR & Nuclear Magnetic Resonance\\
RKKY & Ruderman--Kittel--Kasuya--Yosida\\
BMP & Bound Magnetic Polaron\\
FC & Field Cooling\\
ZFC & Zero Field Cooling
\end{tabular}}


%=====================================
% References, variant A: internal bibliography
%=====================================
\reftitle{References}

% The following MDPI journals use author-date citation: Arts, Econometrics, Economies, Genealogy, Humanities, IJFS, JRFM, Laws, Religions, Risks, Social Sciences. For those journals, please follow the formatting guidelines on http://www.mdpi.com/authors/references
% To cite two works by the same author: \citeauthor{ref-journal-1a} (\citeyear{ref-journal-1a}, \citeyear{ref-journal-1b}). This produces: Whittaker (1967, 1975)
% To cite two works by the same author with specific pages: \citeauthor{ref-journal-3a} (\citeyear{ref-journal-3a}, p. 328; \citeyear{ref-journal-3b}, p.475). This produces: Wong (1999, p. 328; 2000, p. 475)

%=====================================
% References, variant B: external bibliography
%=====================================
%\externalbibliography{yes}
%\bibliography{your_external_BibTeX_file}

%%%%%%%%%%%%%%%%%%%%%%%%%%%%%%%%%%%%%%%%%%
%% optional
%\sampleavailability{Samples of the compounds......  are available from the authors.}

%% for journal Sci
%\reviewreports{\\
%Reviewer 1 comments and authors’ response\\
%Reviewer 2 comments and authors’ response\\
%Reviewer 3 comments and authors’ response
%}

%%%%%%%%%%%%%%%%%%%%%%%%%%%%%%%%%%%%%%%%%%
\end{document}